\documentclass{elsart}

\usepackage{epsfig}

\begin{document}
\begin{frontmatter}

\title{Mixing in the Solar Nebula: Implications for Isotopic 
Heterogeneity and Large-Scale Transport of Refractory Grains}

\author{Alan P. Boss}

\ead{boss@dtm.ciw.edu}

\ead[url]{http://www.dtm.ciw.edu/boss}

\address{DTM, Carnegie Institution, 5241 Broad Branch Road, N.W.,
Washington, D.C. 20015-1305, U.S.A.}

\begin{abstract}

The discovery of refractory grains amongst the particles collected from
Comet 81P/Wild 2 by the Stardust spacecraft (Brownlee et al. 2006) 
provides the ground truth for large-scale transport of materials
formed in high temperature regions close to the protosun outward to
the comet-forming regions of the solar nebula. While accretion disk
models driven by a generic turbulent viscosity have been invoked as
a means to explain such large-scale transport, the detailed physics
behind such an ``alpha'' viscosity remains unclear. We present here an
alternative physical mechanism for large-scale transport in the solar
nebula: gravitational torques associated with the transient spiral arms
in a marginally gravitationally unstable disk, of the type that appears
to be necessary to form gas giant planets. Three dimensional models
are presented of the time evolution of self-gravitating disks,
including radiative transfer and detailed equations of state, showing
that small dust grains will be transported upstream and downstream
(with respect to the mean inward flow of gas and dust being accreted
by the central protostar) inside the disk on time scales of less than
1000 yr inside 10 AU. These models furthermore show that any initial
spatial heterogeneities present (e.g., in short-lived isotopes such
as $^{26}$Al) will be homogenized by disk mixing down to a level of
$\sim$ 10\%, preserving the use of short-lived isotopes as accurate nebular 
chronometers, while simultaneously allowing for the spread of stable 
oxygen isotope ratios. This finite level of nebular spatial heterogeneity
appears to be related to the coarse mixing achieved by spiral arms,
with radial widths of order 1 AU, over time scales of $\sim$ 1000 yrs.
  
\end{abstract}


\end{frontmatter}

\section{Introduction}

 Comets have long been recognized as relatively pristine repositories of
particles from the early solar system, and combined with primitive
chondritic meteorites, these objects represent the best records we have
of the events in the solar nebula some 4.57 Gyr ago. For this very
reason, the Stardust Mission to Comet 81P/Wild 2 was conceived and
launched to sample the particles in the coma of this Jupiter family
comet (JFC). JFCs are believed to have formed in the Kuiper Belt, just
beyond Neptune's orbit, and so Wild 2 is thought to be a sample of
the outermost planet-forming region of the solar nebula. The discovery 
of refractory grains by Stardust (Brownlee et al. 2006) therefore
provided proof that grains that formed in high temperature regions
of the disk, presumably close to the protosun, were able to be
transported outward tens of AUs to the JFC-forming region prior to
their incorporation in comets (see also Nuth \& Johnson 2006).

 There are two possibilities for the outward transport of the
high temperature minerals found by Stardust. First, refractory grains 
formed close to the protosun may have been carried upward by 
the protosun's bipolar outflow and lofted onto trajectories that would 
return them to the surface of the solar nebula at much greater distances 
(Shu et al. 2001). Bipolar outflows are known to occur, but the
lofting of grains and their return to the outer disk are processes
that are not constrained by astronomical observations, though the
Stardust discoveries can be interpreted as proof of this possibility.
Alternatively, refractory solids might have been transported outward 
to comet-forming distances by mixing processes within the solar nebula, such 
as generic accretion disk turbulence (Gail 2002; Ciesla \& Cuzzi 2006; 
Tscharnuter \& Gail 2007; Ciesla 2007), or the actions of spiral arms in a 
marginally gravitationally unstable nebula (Boss 2004, 2007). The
magnetorotational instability (MRI) in an ionized disk is often
considered to be the physical mechanism responsible for disk 
evolution and the source of the disk's effective ``alpha'' viscosity. 
However, MRI effects are limited to the disk's surfaces and its
innermost and outermost regions by the need for appreciable 
fractional ionization. The ionized surface layers are thought to 
be dominated by the inward flow of gas accreting onto the protostar.
The disk's midplane is essentially neutral in the planet-forming 
region from $\sim 1$ to $\sim 15$ AU, preventing the 
MRI from serving as a driver of midplane disk evolution 
in this key region (e.g., Matsumura \& Pudritz 2006). 
Gravitational torques in a marginally gravitationally unstable
disk, however, are able to drive global angular momentum transport
throughout the dead zone, and in fact periods of gravitational
instability may well be regulated by the pile-up of disk gas on
the edge of the dead zone by the action of the MRI in the more
ionized regions of the disk.

  Astronomical observations of disks around young stars often find 
evidence for crystalline silicate grains at distances ranging from 
inside 3 AU to beyond 5 AU, in both the disk's midplane and its surface 
layers (Mer\'in et al. 2007). These grains could have been produced 
through thermal annealing of amorphous grains by hot disk temperatures
reached only within the innermost disk, well inside 1 AU. Crystalline and 
amorphous silicate grains thus appear to require large-scale transport as well.
In addition, isotopic evidence suggests (e.g., Bizzarro, Baker \& Haack 2004) 
that some Allende chondrules formed with an $^{26}$Al/$^{27}$Al ratio 
similar to that of calcium, aluminum-rich inclusions (CAIs). 
Chondrule formation thus appears to have begun shortly after CAI 
formation, and to have lasted for $\sim$ 1 Myr to 4 Myr (Amelin et al. 2002). 
While chondrules are generally believed to have been melted by
shock-front heating at asteroidal distances (e.g., Desch \& Connolly 2002), 
CAIs are thought to have formed much closer to the protosun, again because 
of their refractory compositions. Transport of solids from the inner
solar nebula out to asteroidal distances thus seems to be required in order
to assemble chondrules, CAIs, and matrix grains into the chondritic 
meteorites (Boss \& Durisen 2005). 

 While the evidence for large-scale transport is clear, the degree of 
mixing of the nebula gas and solid has been less clear for some time.
Isotopic evidence has been presented for both homogeneity (e.g., Hsu, 
Huss \& Wasserburg 2003) and heterogeneity (e.g., Simon et al. 1998) 
of short-lived radioisotopes (SLRI) such as $^{26}$Al and $^{60}$Fe in the 
solar nebula. The solar system's $^{60}$Fe must have been synthesized in 
a supernova (Mostefaoui, Lugmair \& Hoppe 2005; Tachibana et al. 2006) 
and then injected either into the presolar cloud (Vanhala \& Boss 2002) 
or onto the surface of the solar nebula (e.g., Boss 2007), 
processes that both lead to initially strongly heterogeneous distributions. 
The initial degree of homogeneity of the SLRI is crucial if $^{26}$Al/$^{27}$Al 
ratios are to be used as precise chronometers for the early solar 
system (Bizzarro et al. 2004; Halliday 2004; Krot et al. 2005; Thrane
et al. 2006). 

 The degree of uniformity is equally murky even for stable isotopes.
For samarium and neodymium isotopes, both nebular homogeneity and 
heterogeneity have been claimed, depending on the objects
under consideration (Andreasen \& Sharma 2007), e.g., ordinary
versus carbonaceous chondrites. However, osmium isotopes seem to require
a high degree of homogeneity in both ordinary and carbonaceous 
chondrites (Yokoyama et al. 2007). The three stable isotopes of 
oxygen, on the other hand, show clear evidence for heterogeneity in 
primitive meteorites (Clayton 1993). This heterogeneity is perhaps 
best explained by self-shielding of molecular CO gas from 
UV photodissociation at the surface of the solar nebula (Clayton 2002;
Lyons \& Young 2005; however, see Ali \& Nuth 2007 for an opposing view) 
or in the presolar molecular cloud (Yurimoto \& Kuramoto 2004).
Oxygen isotopic anomalies formed at the outer surfaces of the solar nebula 
by definition imply a spatially heterogeneous initial distribution,
perhaps capable of explaining the entire range of stable oxygen
isotope ratios through mixing between $^{16}$O-rich and 
$^{16}$O-poor reservoirs (e.g., Sakamoto et al., 2007).
 
 We present here a new set of three dimensional hydrodynamical models
of mixing and transport in a protoplanetary disk that demonstrate how
the required large-scale transport and mixing may have occurred in 
the solar nebula. These models are similar to those of Boss (2004, 2007) 
except that in the new models, the disk is assumed to extend
from 1 AU to 10 AU, instead of from 4 AU to 20 AU, in order to better 
represent the regions of the nebula of most interest for high temperature
thermal processing and meteorite formation.

\section{Numerical Methods}

 The calculations were performed with a numerical code that uses 
finite differences to solve the three dimensional equations of 
hydrodynamics, radiative transfer, and the Poisson equation for the
gravitational potential. Detailed equations of state for the
gas (primarily molecular hydrogen) and dust grain opacities are
employed in the models. Radiative transfer is handled in the
diffusion approximation, which is valid near the disk midplane
and throughout most of the disk, because of the high vertical optical
depths. Artificial viscosity is not employed. A color equation is
solved in order to follow the evolution of initially spatially heterogeneous
tracers, such as SLRI. The color equation is identical to the density
equation, except that the color field is a passive field (a dye) and does
not act back on the evolution of the disk. The energy equation is solved
explicitly in conservation law form, as are the five other hydrodynamic
equations. The code is the same as that used in previous studies 
of mixing and transport in disks (Boss 2004, 2007), and has been shown to
be second-order-accurate in both space and time through convergence testing
(Boss \& Myhill 1992). 

 The equations are solved on a three dimensional, spherical coordinate
grid. The number of grid points in each spatial direction is: $N_r = 51$,
$N_\theta = 23$ in $\pi/2 \ge \theta \ge 0$, and $N_\phi = 256$. The radial 
grid is uniformly spaced between 1 and 10 AU, with boundary conditions 
at both the inner and outer edges chosen to absorb radial velocity
perturbations. The $\theta$ grid is compressed into the midplane to ensure
adequate vertical resolution ($\Delta \theta = 0.3^o$ at the midplane).
The $\phi$ grid is uniformly spaced, to prevent any bias in the azimuthal
direction. The central protostar wobbles in response to the growth of
nonaxisymmetry in the disk, thereby rigorously preserving the location of
the center of mass of the star and disk system. The number of terms in the
spherical harmonic expansion for the gravitational potential of the disk
is $N_{Ylm} = 32$.

 In some of the models, diffusion of the dust grains carrying the 
SLRI or oxygen isotope anomalies with respect to the disk gas is 
modeled through a modification of the color equation (Boss 2004, 2007). 
This modification consists of adding a diffusion term (second space 
derivative of the color field), multiplied by an appropriate 
diffusion coefficient $D$, which is approximated by the eddy viscosity 
of a classical viscous accretion disk: $D = \alpha h c_s$, 
where $\alpha =$ disk turbulent viscosity parameter, $h =$ disk
scale height, and $c_s =$ isothermal sound speed at the disk midplane.
For typical nebula values at 5 to 10 AU ($h \approx 1$ AU, $T 
\approx 100$ K, $c_s \approx 6 \times 10^4$ cm s$^{-1}$), the eddy
diffusivity becomes $D = 10^{18} \alpha$ cm$^2$ s$^{-1}$.

\section{Initial Conditions}

 The models consist of a $1 M_\odot$ central protostar surrounded
by a protoplanetary disk with a mass of 0.047 $M_\odot$ between 1 and
10 AU. The underlying disk structure is the same as that of the disk
extending from 4 AU to 20 AU in the models of Boss (2004, 2007). 
Disks with similar masses and surface densities appear to 
be necessary to form gas giant planets by either core accretion (e.g.,
Inaba et al. 2003) or by disk instability (e.g., Boss et al. 2002).
Chemical species observed in comets set an upper limit for disk 
midplane temperatures in the outer disk of $\sim$ 50 K (Boss 1998).
The combination of a disk massive enough to form gas giant planets
and the maximum temperatures consistent with cometary speciation
implies that the solar nebula was at least marginally gravitationally
unstable. Astronomical observations have long indicated that
protoplanetary disks have masses in the range of 0.01 to 0.1 $M_\odot$,
(e.g., Kitamura et al. 2002), but only more recently has it become
apparent that these disk masses appear to be underestimated by
factors of up to 10 (Andrews \& Williams 2007), which is reassuring,
given the high frequency of extrasolar gas giant planets and the need
for sufficiently massive disks to explain their formation.

 The disks start with an outer disk temperature $T_o = 40$ K, leading
to a minimum in the Toomre $Q$ ($Q = \Omega c_s / \pi G \sigma$,
where $\Omega$ is the disk angular velocity, $c_s \propto T^{1/2}$ 
is the sound speed, $G$ is the gravitational constant, and $\sigma$ is 
the disk gas surface density) value of 1.43 at 10 AU; 
inside 6.5 AU, $Q$ rises to higher values because
of the much higher disk temperatures closer to the protosun.
A $Q$ value of 1.43 implies marginal instability to the growth
of gravitationally-driven perturbations, while high values of $Q$ 
imply a high degree of gravitational stability. These disk models therefore
represent marginally gravitationally unstable disks, where strong
spiral arms are expected to form on a dynamical (rotational) time
scale, and dominate the subsequent evolution of the disk. 

 A color field representing SLRI or oxygen anomalies is sprayed onto 
the outer surface of the disk (or injected into the midplane in one
model) at a radial distance of either 2.2 or 8.6 AU, into a 90 degree 
(in the azimuthal direction) sector of a ring of width 1 AU, simulating 
the arrival of a Rayleigh-Taylor finger of supernova injecta
(e.g., Vanhala \& Boss 2002), a spray of hot grains lofted by
the X-wind bipolar outflow (Shu et al. 2001), or a region of the
disk's surface that has undergone oxygen isotope fractionation. One 
of the models includes the effects of the diffusion of the color field with 
respect to the gaseous disk by a generic turbulent viscosity 
characterized by $\alpha = 0.0001$. This low $\alpha$ value 
is small enough to have only a very minor effect on the color field 
(Boss 2004, 2007). SLRI, refractory grains, or oxygen anomalies that reside 
in the gas or in particles with sizes of mm to cm or smaller will remain 
effectively tied to the gas over timescales of $\sim$ 1000 yrs or so, 
because the relative motions caused by gas drag result in 
differential migration by distances of less than 0.1 AU in 1000 yrs, 
which is negligible compared to the distances they are transported 
by the gas in that time, justifying their representation by the
color field. 

\section{Results}

 We present here the results of four models. Two models represent
initial heterogeneity of the color field on the disk surface at 
distances of either 2 AU (model 2S) or 9 AU (model 9S) with $\alpha = 0$,
as well as a third model with 9 AU surface injection with nonzero diffusivity 
$\alpha  = 0.0001$ (model 9SD). The fourth model assumes the
initial heterogeneity exists in the disk's midplane at 2 AU (model
2M), as would be appropriate for refractory grains heated by
the hot inner disk. We shall see that whether or not $\alpha = 0$
or 0.0001 makes little difference to the outcome, as the evolution
is dominated by the actions of the gravitational torques from the spiral
arms that form. Similarly, whether the color field starts at the disk's 
surface or midplane makes little difference, as the vertical convective-like
motions in these disks rapidly mix the color field upward and downward
on a time scale comparable to the orbital period (Boss 2004, 2007).

 The color fields shown in the figures represent the number density
of small solids (e.g., chondrules, CAIs, ice grains, or their precursor 
grain aggregates) in the disk carrying SLRI, refractory minerals, or 
oxygen isotope anomalies, which evolve along with the disk's gas.

 Fig. 1 shows how the initial color fields are limited to 90 degree
arcs in either the midplane or on the surface of the disks. Fig. 2
shows the midplane color field for model 2S after 385 yrs. Even
though the color was initially sprayed onto the surface of model
2S at an orbital distance of 2 AU, the color is rapidly transported 
down to the disk's midplane by convective-like motions (Boss 2007).
In addition, the net effect of the evolution of this marginally 
gravitationally unstable disk is to transport the color field inward 
and outward to the disk boundaries at 1 AU and 10 AU in less than 385 yr. 
The color field in the innermost disk is depleted by its accretion
onto the central protostar, while the color that is transported outward
is forced to pile up at the outer disk boundary. This model
demonstrates clearly the rapid outward (and upstream with respect
to the inward accretion flow of disk gas) transport of small
grains from the hot inner disk to the cooler outer disk regions,
on time scales of less than 1000 yr.
 
 The color field shown in Fig. 2 represents, e.g., the number density
of SLRIs in a disk after $\sim 385$ yr of evolution following
an injection event. Given the strong gradients in the color
fields evident in Fig. 2, it is clear that the color field is highly 
spatially heterogeneous. This is because the underlying gas density
distribution is equally highly non-axisymmetric and the gas density 
is the mechanism that drives the transport of the color field.
However, cosmochemical SLRI abundances typically are given as 
ratios, i.e., $^{26}$Al/$^{27}$Al, where the SLRI $^{26}$Al is derived 
from the injection event, while the reference stable isotope $^{27}$Al 
is presumably derived primarily from a well-mixed presolar cloud
and so is homogeneous with respect to the original gas disk.
[Additional $^{27}$Al would be expected to accompany any $^{26}$Al 
derived from a supernova source, though not from an X-wind source.]
In order to determine the variations in the SLRI ratio
$^{26}$Al/$^{27}$Al, then, the color field must be normalized
by the gas density. This has been done for model 2S in Fig. 3, which
shows the log of the color/gas ratio at a time of 8.4 yr, shortly
after the injection event. The high degree of initial spatial
heterogeneity of the ratio $^{26}$Al/$^{27}$Al is obvious.
However, Fig. 4 shows that after 385 yr of evolution (the same time 
as in Fig. 2), the color/gas ratio has become remarkably uniform 
throughout the midplane of the disk, with variations of
less than 0.1 in the log, or no more than a factor of 1.26.
Even variations this large only occur close to
the inner and outer boundary where there is very little
color and even less gas density, so that the ratio becomes large. 
Such regions contain essentially no gas and so are negligible. 
This model shows that the color field can become homogeneous with 
respect to the disk gas in a few hundred years of evolution
in a marginally gravitationally unstable disk.

 An important quantity to calculate in this regard is the dispersion 
of the color/gas density ratio from its mean value (Boss 2007), in 
order to make a more quantitative comparison with isotopic measurements.
Fig. 5 shows the time evolution of the level of spatial heterogeneity
in, e.g., SLRI ratios such as $^{26}$Al/$^{27}$Al following a single 
SLRI injection event, for models 2S and 2M. Fig. 5 plots the 
square root of the sum of the squares of the color field 
divided by the gas density subtracted from the mean value of the 
color field divided by the gas density, where the sum
is taken over the midplane grid points and is normalized by
the number of grid points being summed over. The sum excludes
the regions closest to the inner and outer boundaries, as well
as regions with disk gas densities less than $10^{-12}$ g cm$^{-3}$,
as the low gas densities in these regions skew the calculations
of the dispersion of the ratio of color to disk gas, and these
regions contain comparatively little gas and dust. 

 Fig. 5 shows that the dispersions for both models 2S and 2M
follow similar time evolutions, starting from high initial
values. While model 2M has not yet been continued as far as model 2S,
model 2S shows that the expectation is that in both models,
the dispersion will fall to a level of 5\% to 10\%, as was
found in the larger-scale (4 AU to 20 AU) disk models of Boss (2007).
 
 We now turn to the models where the color field was initiated
at orbital distances centered on 9 AU. Fig. 6 shows the result
for model 9SD just 16 yr after injection, when the color has
already been transported down to the midplane, as well as inward
to orbital distances of only a few AU. The spiral-like nature of
the color field is a clear indicator of the transport mechanism
responsible for this rapid movement. Fig. 7 shows how the
color/gas density ratio for model 9SD has become considerably
more homogenized after 210 yr of evolution, while Fig. 8 shows
that nearly complete homogenization of the color field is
achieved after 1526 yrs of evolution, with some residual
heterogeneity remaining in the inner disk. Fig. 9 shows
the color/gas density ratio for model 9S, which was identical
to model 9SD, except for having $\alpha = 0$, compared to
$\alpha = 0.0001$ for model 9SD. Evidently this small level
of diffusivity of the color field with respect to the gas
density has little appreciable effect on the evolution of the
color field, as was also found by Boss (2007).

 Finally, Fig. 10 displays the time evolution of the dispersions
for models 9S and 9SD. Both models demonstrate that the
dispersion does not approach zero, as might be expected, but
rather falls to a level on the order of 10\%, where the dispersion
appears to reach a steady state. This implies that gravitational mixing 
is not 100\% efficient and is unable to completely homogenize an
initial spatial heteogeneity. Gravity is a long-range force, 
only able to drive mixing over length scales where large density 
variations occur, i.e., over length scales comparable
to the disk's spiral arms.  These spiral arms tend to have
radial length scales of no less than an AU or so, much larger
than the radial extent of a grid cell (0.18 AU), which appears to
lead to the large-scale granularity at the 10\% level seen in
Fig. 10.

\section{Conclusions}

 The discovery of refractory minerals in Comet Wild 2 by the Stardust
Mission (Brownlee et al. 2006) is consistent with the rapid (less
than 1000 yr), large-scale (10s of AU) outward transport of small
dust grains in a marginally gravitationally unstable disk. Such large-scale 
transport also seems to be required to explain observations of thermally 
processed, crystalline silicates in comets and protoplanetary disks (e.g., 
Nuth \& Johnson 2006; van Boekel et al. 2005; Mer\'in et al. 2007). Marginally 
gravitationally unstable disks have the additional feature of providing 
a robust source of shock fronts capable of thermally processing 
solids into chondrules (Desch \& Connolly 2002; Boss \& Durisen 2005).
Furthermore, the degree of spatial heterogeneity in the solar nebula 
required for the use of the $^{26}$Al/$^{27}$Al chronometer and for the 
survival of the oxygen isotope anomalies is consistent with the 
results of these models of mixing and transport of initially highly spatially 
heterogeneous tracers in a marginally gravitationally unstable disk 
(Boss 2007). Such a disk appears to be required for the formation of 
Jupiter by either the core accretion (Inaba et al. 2003) or disk instability 
(Boss et al. 2002) mechanisms. Marginally gravitationally unstable 
disk models provide a physical mechanism for self-consistent 
calculations of the mixing and transport of dust grains across
the planet-forming region of the solar nebula.

 Mass accretion rates of the disk onto the central protostar are
highly variable in marginally gravitationally unstable models, 
ranging from rates of $\sim 10^{-6} M_\odot$/yr to 
$\sim 10^{-5} M_\odot$/yr, much higher than the typical stellar mass
accretion rate of $\sim 10^{-8} M_\odot$/yr for a classical T Tauri star, 
but comparable to the rates inferred for T Tauri stars during their 
periodic FU Orionis outbursts (Boss 1996). Recent observations suggest 
that most T Tauri disks have lifetimes of no more $\sim$ 1 Myr 
(Cieza et al. 2007). Combined with a disk mass of $\sim 0.1 M_\odot$, 
this suggests a time-averaged mass accretion rate for solar-mass
T Tauri stars of $\sim 10^{-7} M_\odot$/yr. A disk would then
need to experience phases of gravitational instability lasting 
roughly 1\% to 10\% of its lifetime in order to accomplish
the mass transport implied by these estimates. Phases of gravitational 
instability are thus expected to be transient phases, but robust
and frequent enough to achieve the disk mixing and transport processing
described in this paper.

 I thank Sasha Krot and a second referee for their improvements to the paper, 
and Sandy Keiser for her computer wizardry. This research was supported 
in part by the NASA Planetary Geology and Geophysics Program under grant 
NNG05GH30G and by the NASA Origins of Solar Systems Program under grant 
NNG05GI10G, and is contributed in part to the NASA Astrobiology Institute 
under grant NCC2-1056. Calculations were performed on the Carnegie Alpha 
Cluster, which was supported in part by NSF MRI grant AST-9976645.

\vfill\eject
\centerline{\psfig{figure=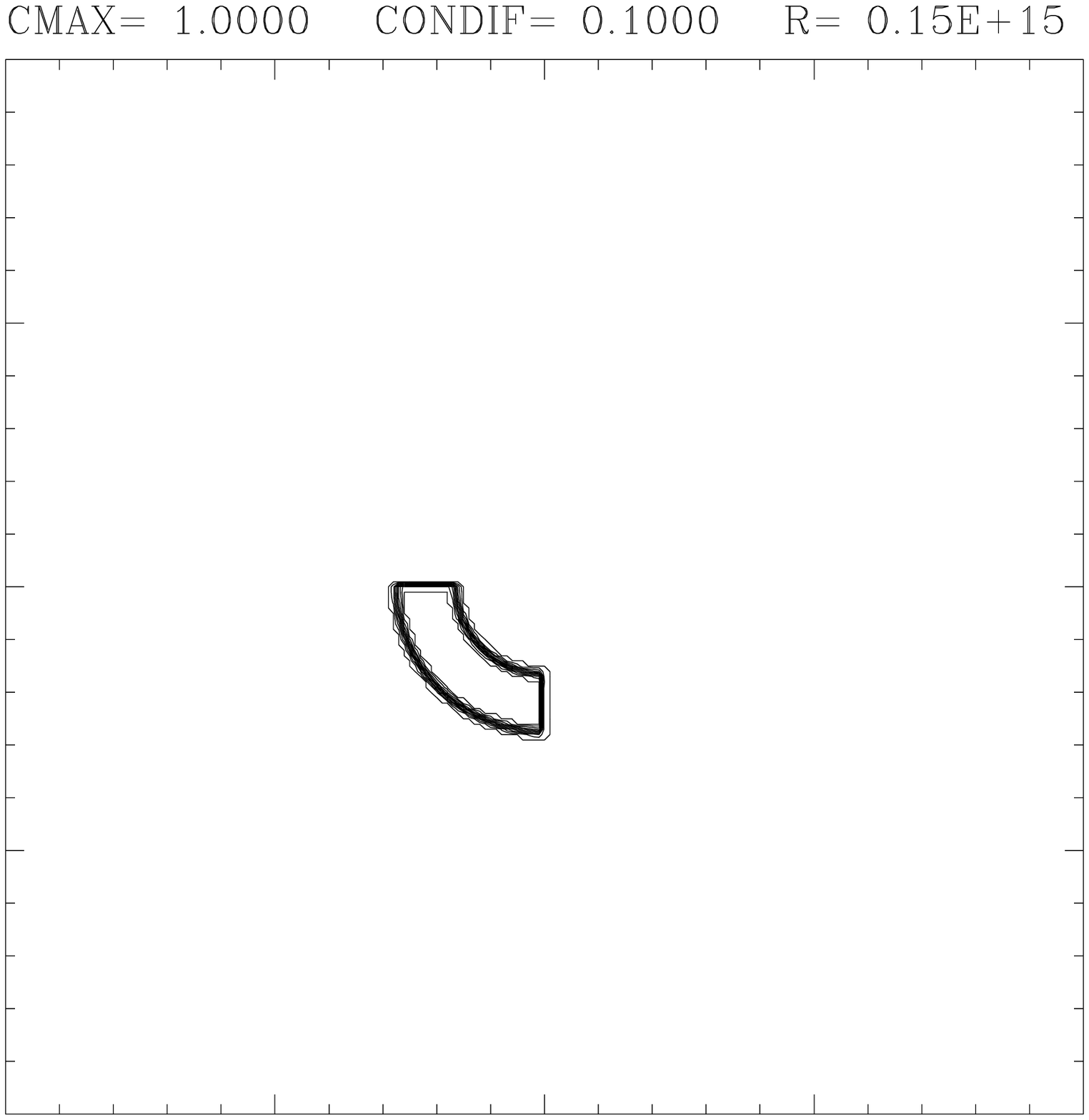,height=8.5in,width=6.2in}}
\vspace{-2.3in}
Figure 1. Model 2M at 0 yr, showing linear contours of 
the color field density (e.g., number of atoms of $^{26}$Al cm$^{-3}$) 
in the disk midplane at the instant when the color field was injected into 
the disk's midplane in a 90 degree azimuthal sector between 1.6 and 
2.8 AU. Region shown is 10 AU in radius (R) with a 1 AU radius inner 
boundary. In this Figure, the contours represent changes in the color 
field density by 0.1 units (CONDIF) on a scale normalized by the 
initial color field density of 1.0, up to a maximum value of 1.0 (CMAX).
\vspace{0.2in}

\vfill\eject
\centerline{\psfig{figure=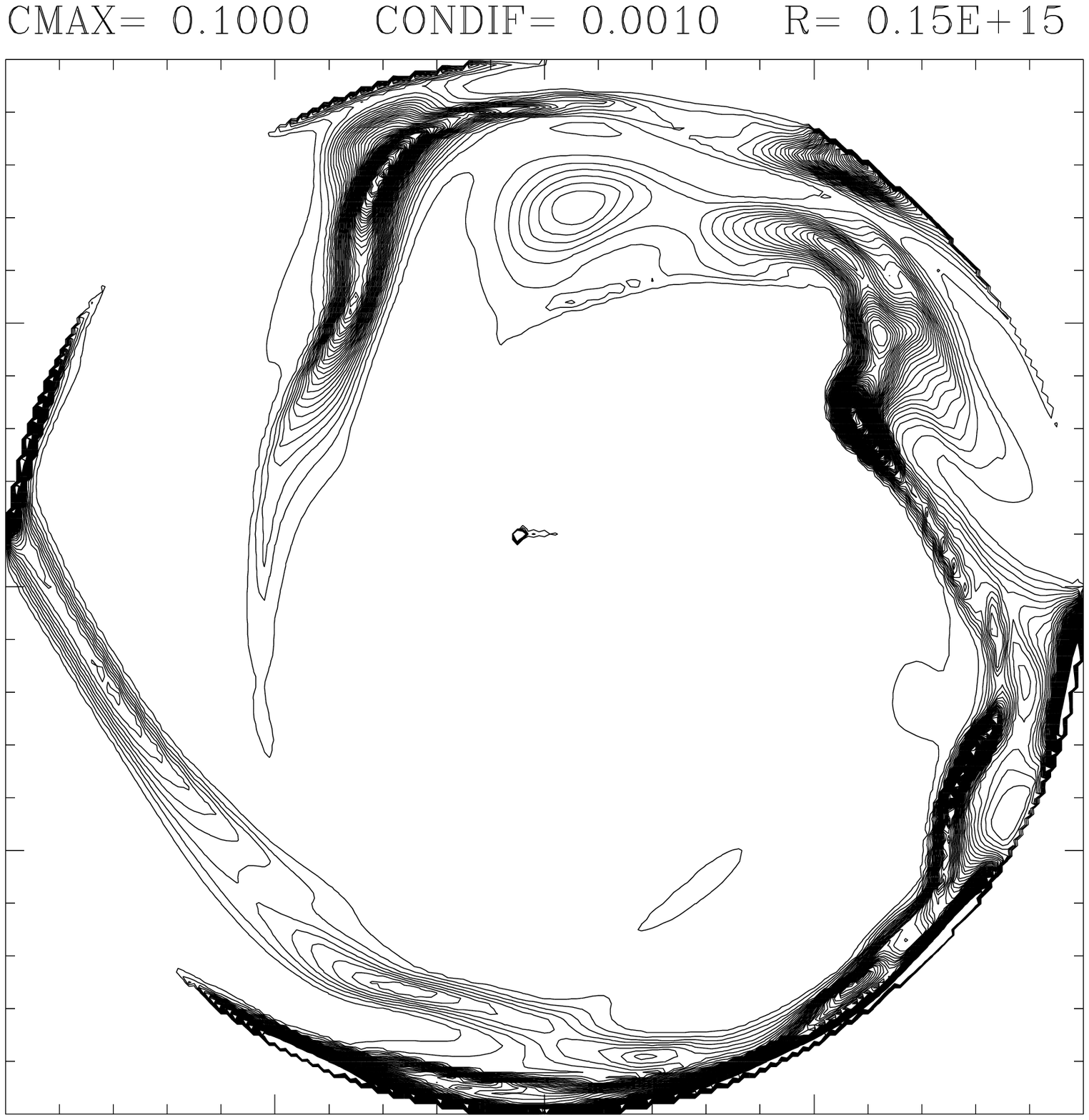,height=8.5in,width=6.2in}}
\vspace{-2.3in}
\noindent
Figure 2. Same as Fig. 1, but for model 2S after 385 yr. The color field 
has spread outward to the edge of the disk. The spatial distribution is 
highly heterogeneous, with the highest concentrations residing inside the 
spiral arms of the disk and on the edge of the disk, where it is 
artificially not allowed to move farther outward.
\vspace{0.2in}

\vfill\eject
\centerline{\psfig{figure=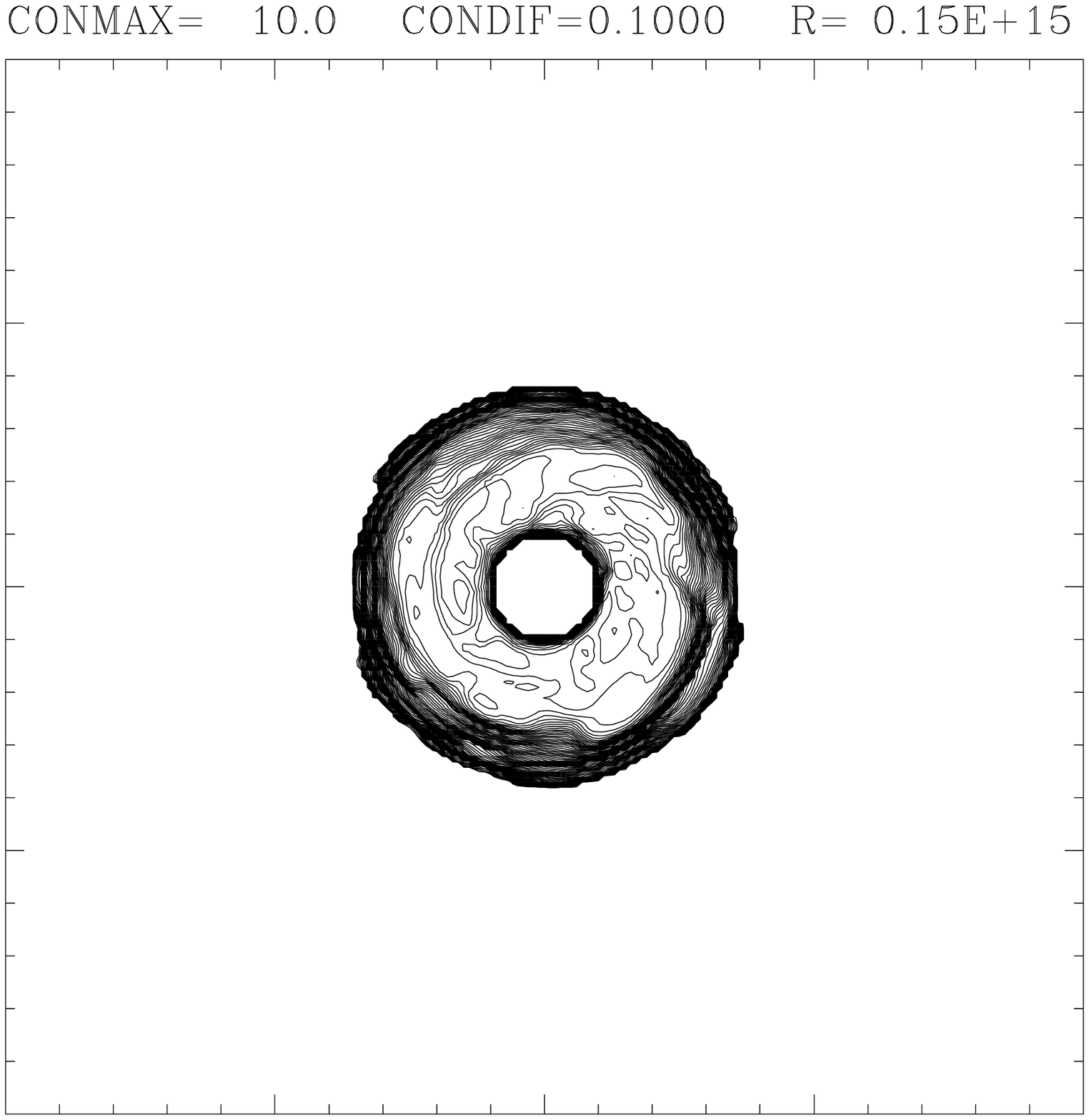,height=8.5in,width=6.2in}}
\vspace{-2.3in}
\noindent
Figure 3. Logarithmic contours of the color field density divided by 
the disk gas density (i.e., log of the abundance ratio $^{26}$Al/$^{27}$Al) 
for model 2S at a time of 8.4 yr, plotted in the same manner as in Fig. 1.
Contours represent changes by factors of 1.26 up to a 
maximum value of 10.0, on a scale defined by the gas disk density.
At this early time, the color/gas ratio is highly spatially
heterogeneous, as the color field is still confined to the
innermost disk, where it was first injected.
\vspace{0.2in}

\vfill\eject
\centerline{\psfig{figure=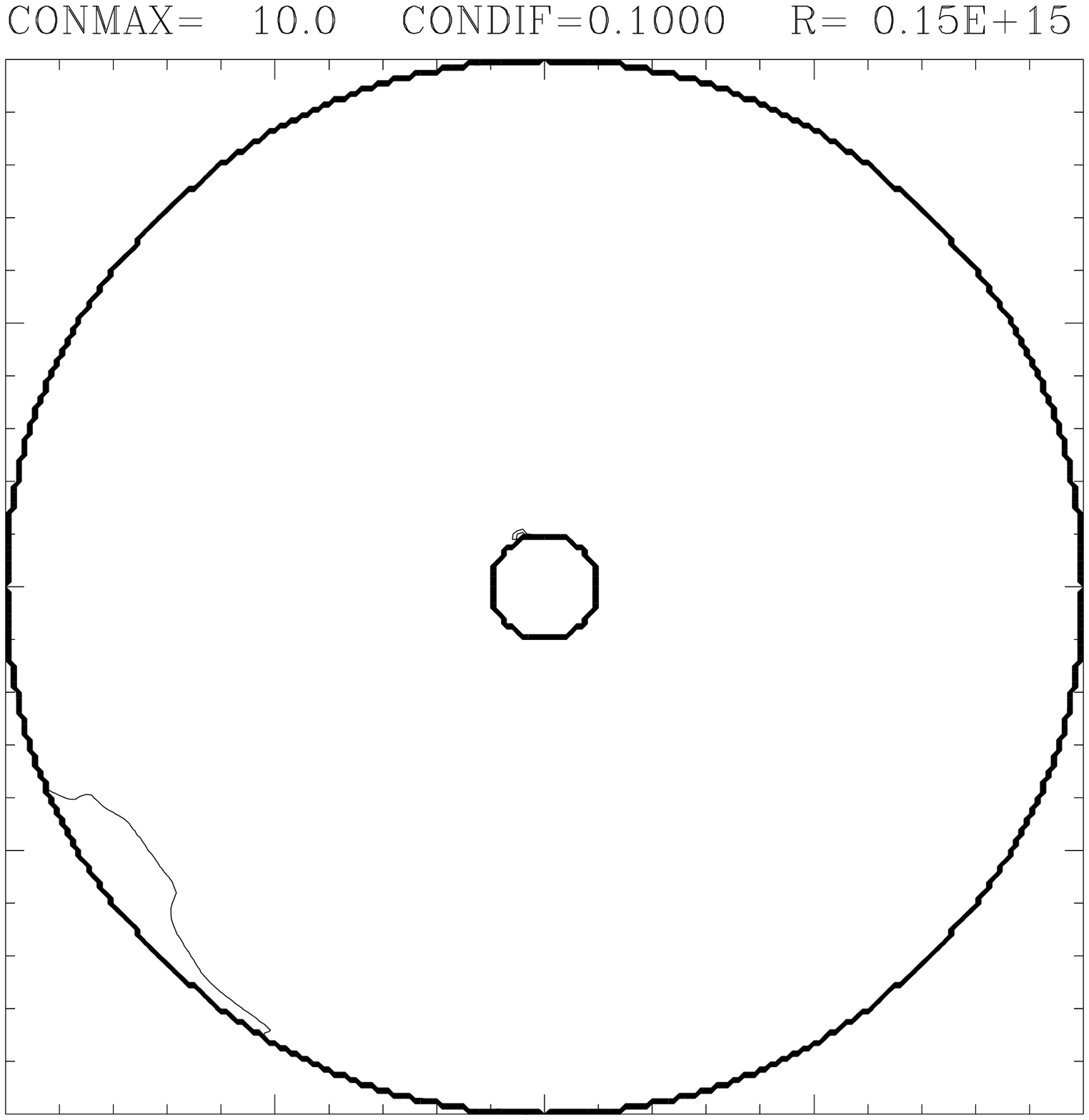,height=8.5in,width=6.2in}}
\vspace{-2.3in}
\noindent
Figure 4. Same as Fig. 3 for model 2S, but after 385 yr of evolution.
The color/gas ratio is now nearly homogeneous throughout the entire
disk, except for small regions along the inner and outer boundaries,
which are clearly artificial boundary effects.

\vspace{0.2in}

\vfill\eject
\centerline{\psfig{figure=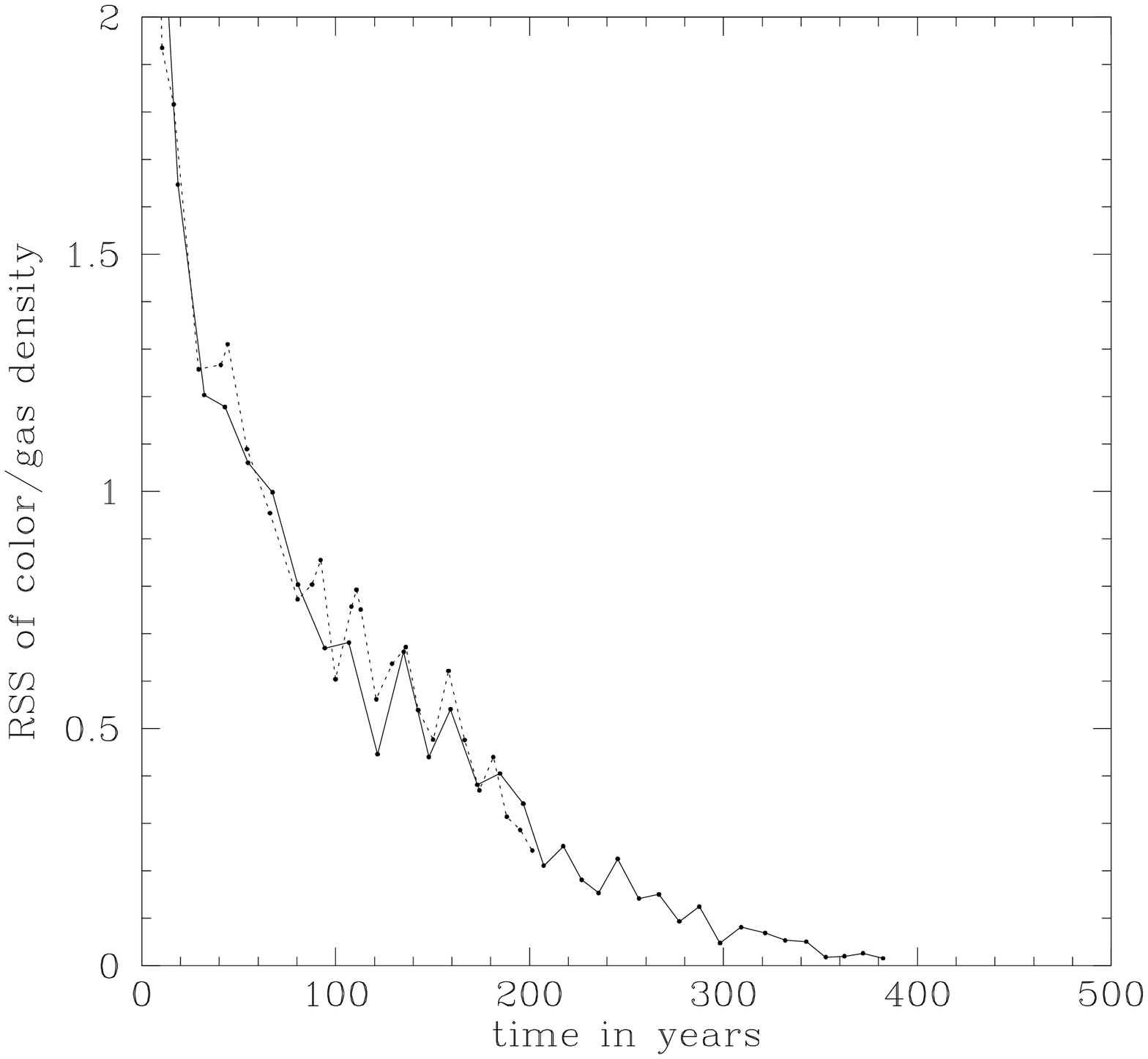,height=8.5in,width=6.2in}}
\vspace{-2.3in}
\noindent
Figure 5. Time evolution of the dispersion from the mean (i.e., standard
deviation, or the root of the sum of the squares [RSS]of the differences
from the mean) of the color field density divided by the gas density
(e.g., $^{26}$Al/$^{27}$Al abundance ratio) in the disk midplane in models 2S 
(solid) and 2M (dashed). The color field is sprayed onto the disk surface 
or injected into the disk midplane at a time of 0 yr. Starting from high 
values (RSS shortly after 0 yrs is $>>$ 1), the dispersion decreases on a 
timescale of $\sim 300$ yrs, dropping to a value of $\sim$ 5\%.
\vspace{0.2in}

\vfill\eject
\centerline{\psfig{figure=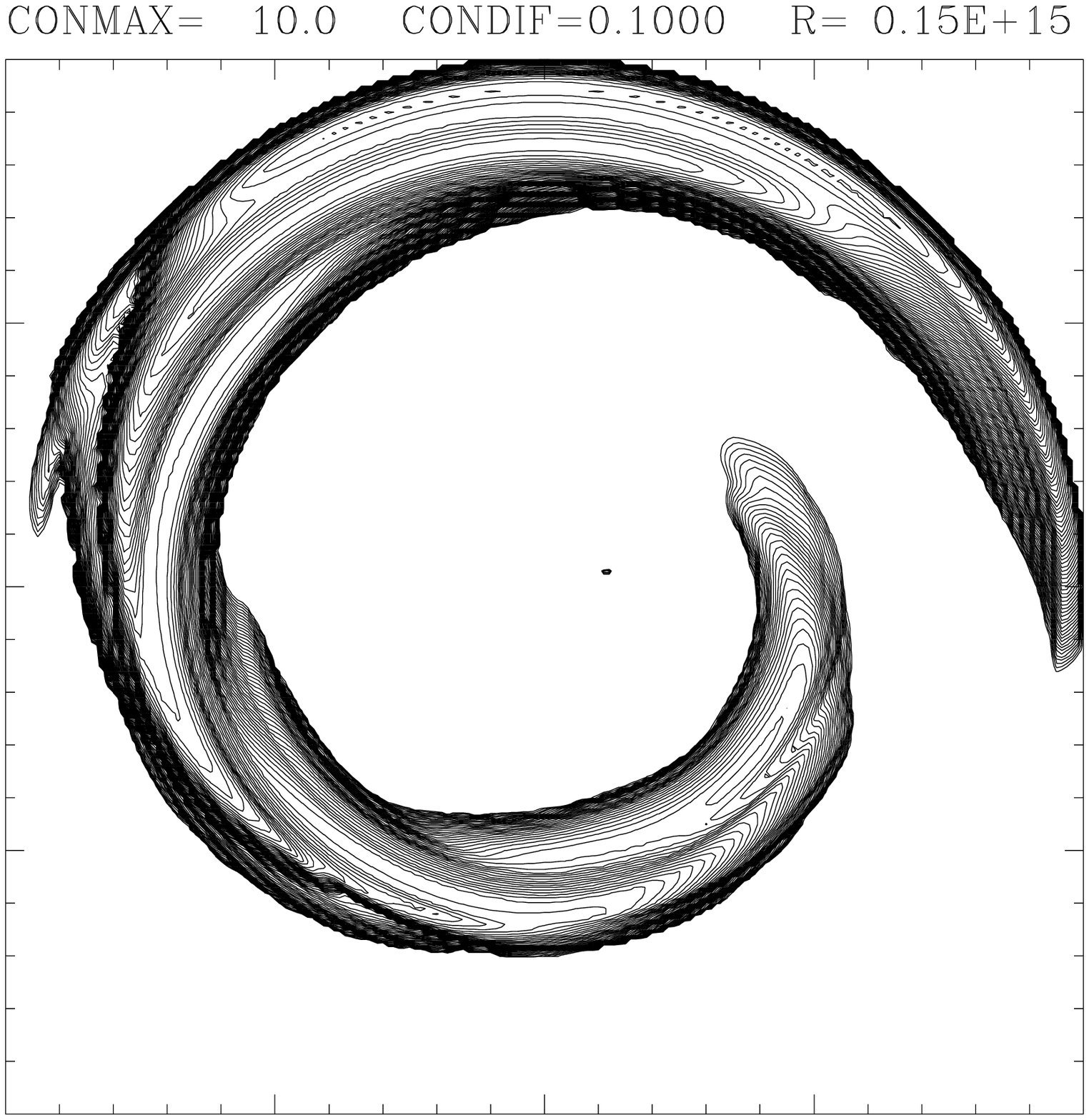,height=8.5in,width=6.2in}}
\vspace{-2.3in}
\noindent
Figure 6. Model 9SD at 16 yr, showing logarithmic contours of the color field
density divided by the disk gas  
density in the disk midplane 16 yr after the color field was sprayed onto
the disk's surface in a 90 degree azimuthal sector between 8.1 and 9.1 AU 
(as in Fig. 1). Region shown is 10 AU in radius (R) with a 1 AU radius inner 
boundary. Contours represent changes by factors of 1.26 up to a 
maximum value of 10.0, on a scale defined by the gas disk density.
At this early time, the color/gas ratio is again highly spatially
heterogeneous, as the color field is mostly confined to the
outermost disk, where it was first injected, though inward motion
toward the central protostar along a single-arm spiral arm is evident.
\vspace{0.2in}

\vfill\eject
\centerline{\psfig{figure=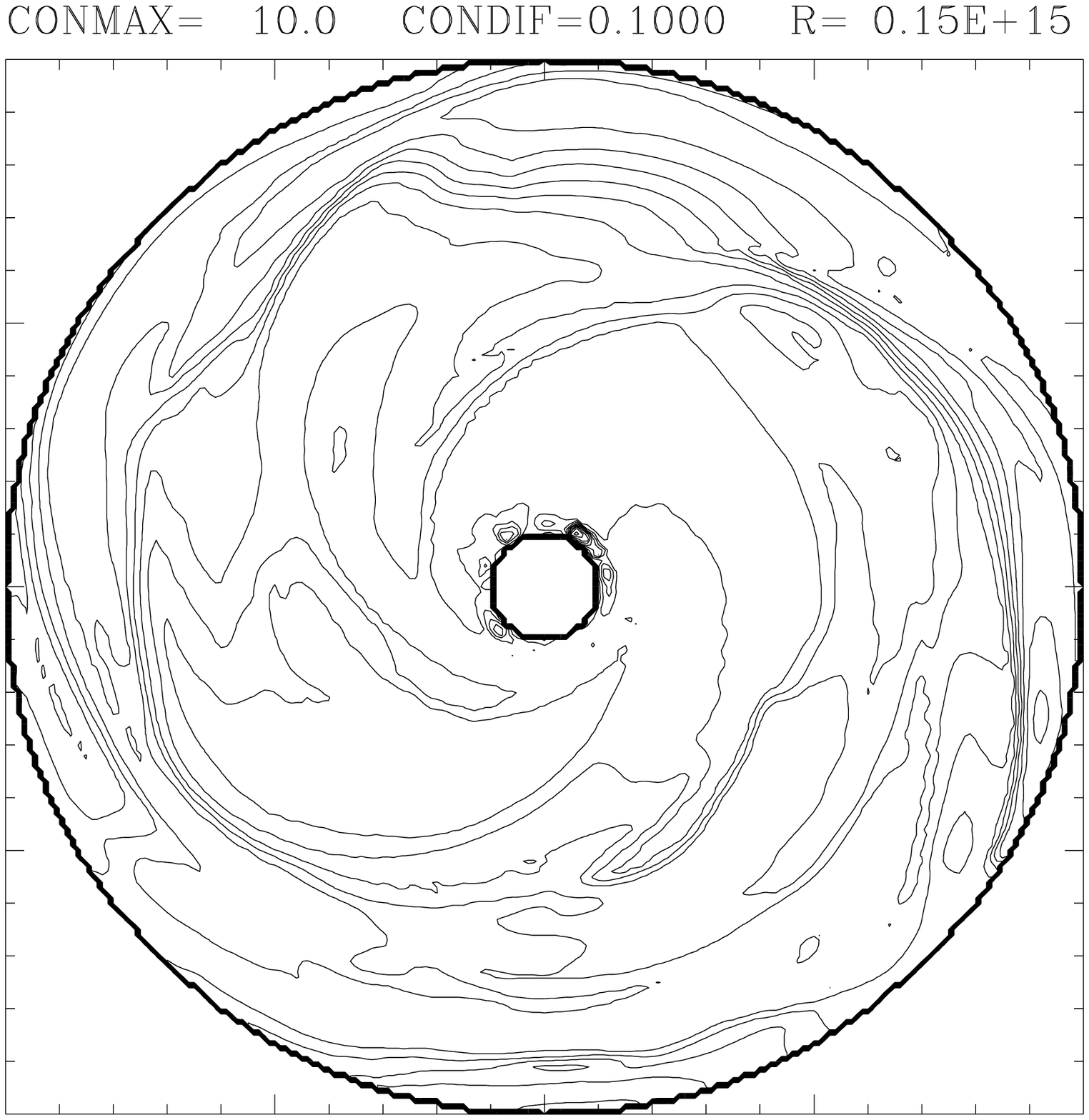,height=8.5in,width=6.2in}}
\vspace{-2.3in}
\noindent
Figure 7. Same as Fig. 6 for model 9SD, but after 210 yr of evolution. 
The color/gas ratio is approaching spatial homogeneity compared to Fig. 6, 
as a result of mixing and transport by the disk's spiral arms.
\vspace{0.2in}

\vfill\eject
\centerline{\psfig{figure=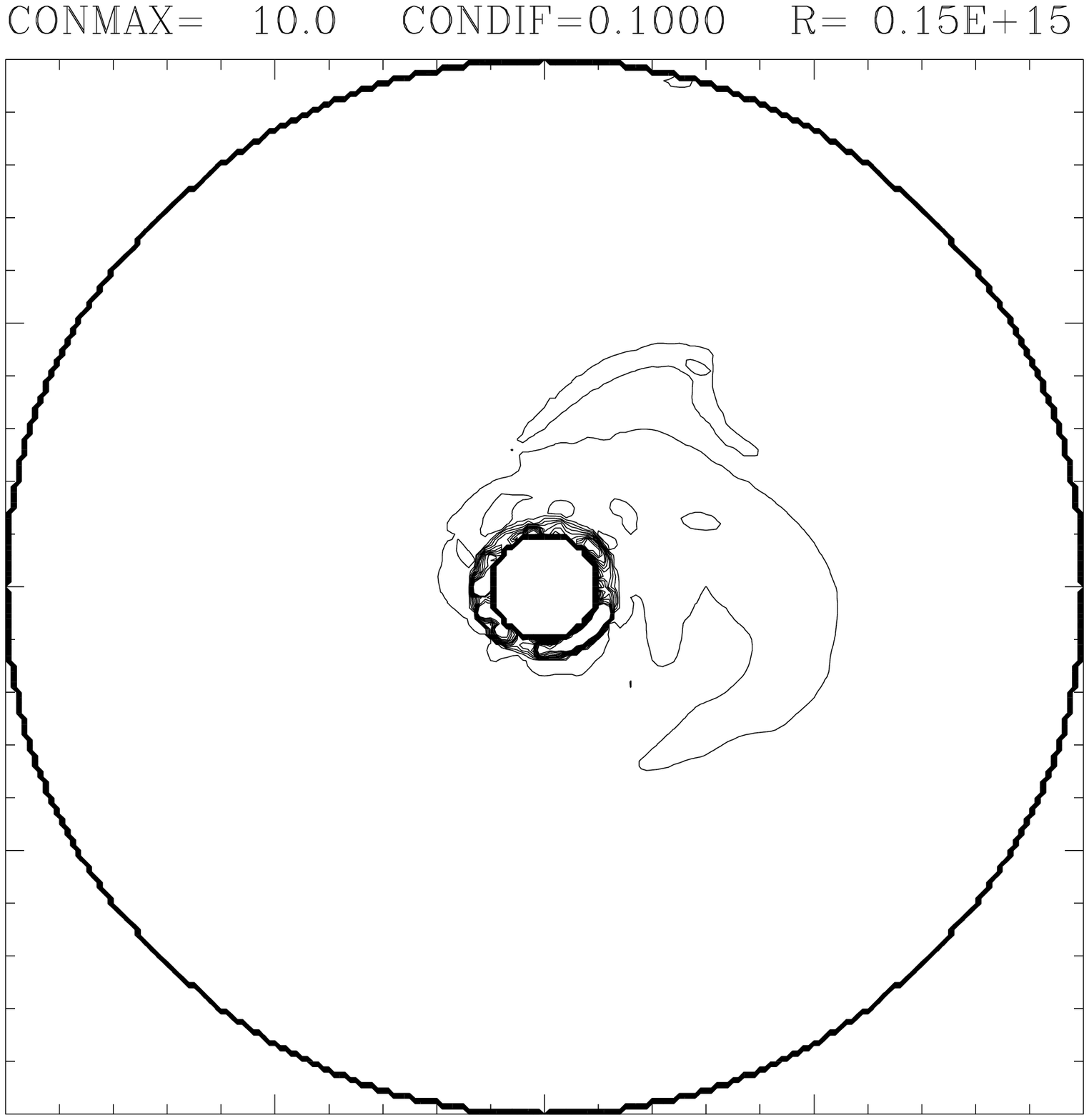,height=8.5in,width=6.2in}}
\vspace{-2.3in}
\noindent
Figure 8. Same as Fig. 7 for model 9SD, but after 1526 yr of evolution.
The color/gas ratio is now nearly homogeneous throughout the entire
disk, except for small regions right along the inner boundary,
which is a boundary effect.
\vspace{0.2in}

\vfill\eject
\centerline{\psfig{figure=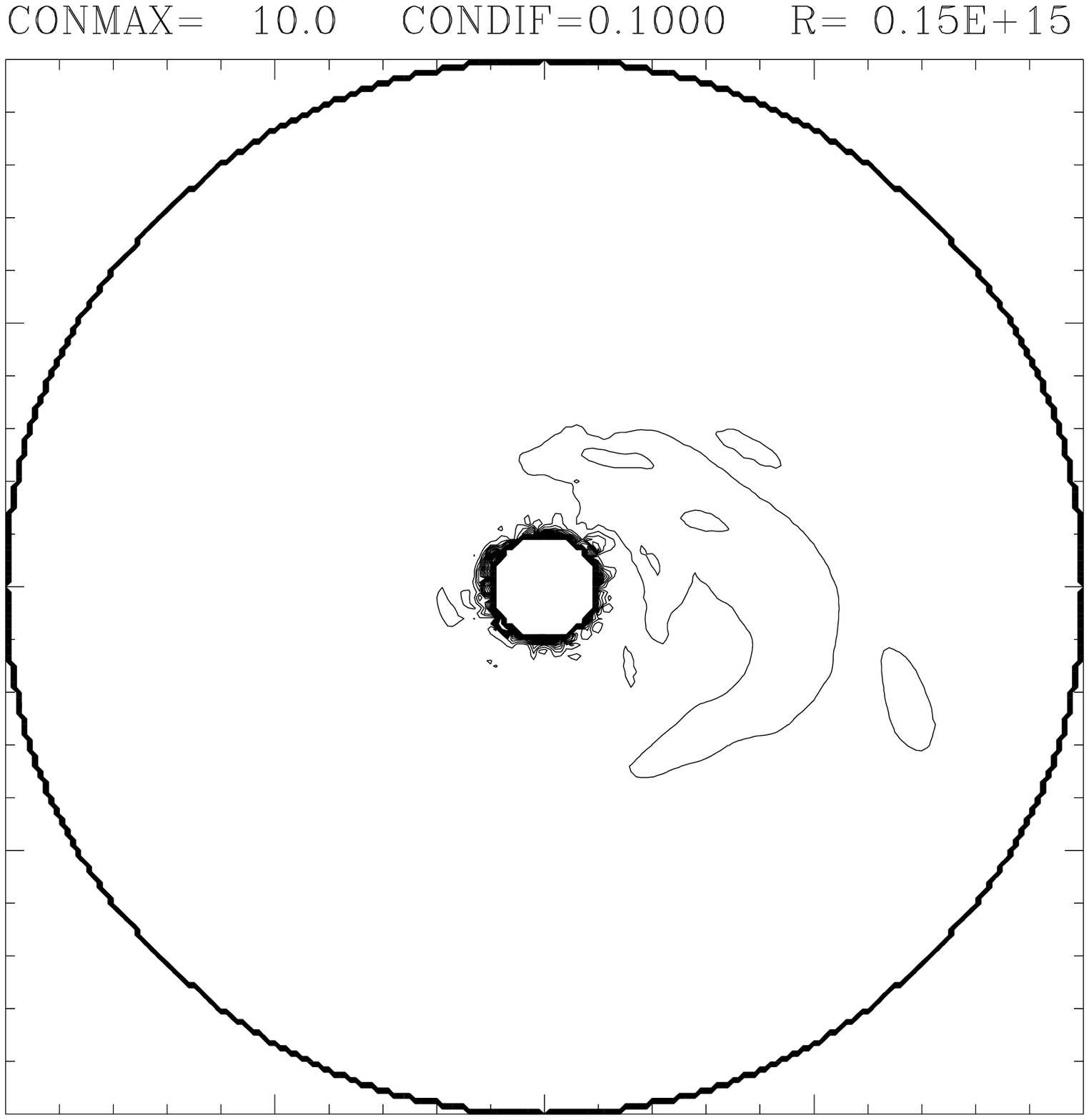,height=8.5in,width=6.2in}}
\vspace{-2.3in}
\noindent
Figure 9. Same as Fig. 8, after 1526 yr of evolution, but for
model 9S, which had no turbulent diffusion of the color field
with respect to the disk gas. Evidently the level of turbulent diffusion
in model 9SD has essentially no effect upon the mixing and transport
processes.
\vspace{0.2in}

\vfill\eject
\centerline{\psfig{figure=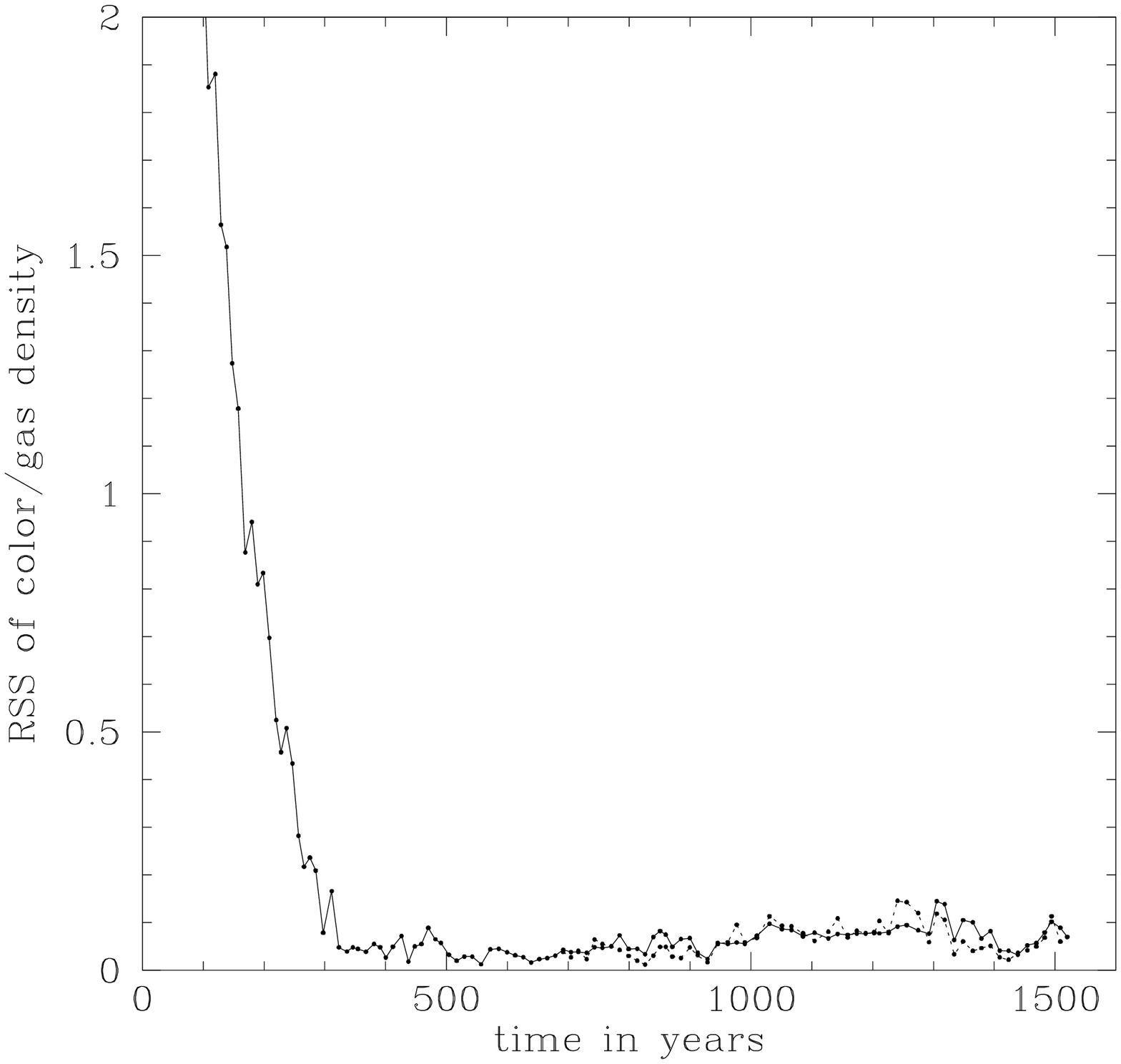,height=8.5in,width=6.2in}}
\vspace{-2.3in}
\noindent
Figure 10. Time evolution of the dispersion from the mean of the color 
field density divided by the gas density (e.g., $^{26}$Al/$^{27}$Al 
abundance ratio) in the disk midplane in models 9SD (solid) and 9S 
(dashed). The color field is sprayed onto the disk surface at a time 
of 0 yr. The dispersion again decreases on a timescale of $\sim 300$ yrs, 
then approaches a steady state value of $\sim$ 10\%. After $\sim$ 700 yr,
the results for model 9S are plotted as well, showing that the non-zero
turbulent diffusion in 9SD has no appeciable effect on the steady level
reached.

\vspace{0.2in}


\begin{thebibliography}{}

\bibitem{} 
Ali, A., \& Nuth, J. A., The oxygen isotope effect in the earliest 
processed solids in the solar system: is it a chemical mass-independent
process?, 2007, Astronomy \& Astrophysics, 467, 919-923.

\bibitem{} 
Amelin, Y., Krot, A. N., Hutcheon, I. D., \& Ulyanov, A. A., 
Lead Isotopic Ages of Chondrules and Calcium-Aluminum-Rich Inclusions,
2002, Science, 297, 1678-1683.

\bibitem{} 
Andreasen, R., \& Sharma, M., Mixing and Homogenization in the Early
Solar System: Clues from Sr, Ba, Nd, and Sm Isotopes in Meteorites,
2007, Astrophysical Journal, 665, 874-883.

\bibitem{} 
Andrews, S. M., \& Williams, J. P., High-Resolution Submillimeter
Constraints on Circumstellar Disk Structure, Astrophysical Journal,
659, 705-728.

\bibitem{} 
Bizzarro, M., Baker, J. A., \& Haack, H., Mg isotope evidence for
contemporaneous formation of chondrules and refractory inclusions, 2004,
Nature, 431, 275-278.

\bibitem{} 
Boss, A. P., Evolution of the Solar Nebula. III. Protoplanetary Disks
Undergoing Mass Accretion, 1996, Astrophysical Journal, 469, 906-920.

\bibitem{} 
Boss, A. P., Temperatures in Protoplanetary Disks, 1998, Annual
Reviews of Earth and Planetary Science, 26, 53-80.

\bibitem{} 
Boss, A. P., Evolution of the Solar Nebula. VI. Mixing and Transport
of Isotopic Heterogeneity, 2004, Astrophysical Journal, 616, 1265-1277.

\bibitem{}
Boss, A. P., Evolution of the Solar Nebula. VIII. Spatial and Temporal
Heterogeneity of Short-lived Radioisotopes and Stable Oxygen Isotopes, 
2007, Astrophysical Journal, 660,  1707-1714.

\bibitem{} 
Boss, A. P., \& Durisen, R. H., Chondrule-Forming Shock Fronts in
the Solar Nebula,  A Possible Unified Scenario for Planet and Chondrite
Formation, 2005, Astrophysical Journal, 621, L137-L140.

\bibitem{} 
Boss, A. P., \& Myhill, E. A., Protostellar Hydrodynamics,  
Constructing and Testing a Spatially and Temporally Second Order Accurate 
Method. I. Spherical Coordinates, 1992, 
Astrophysical Journal Supplement Series, 83, 311-327.

\bibitem{} 
Boss, A. P., Wetherill, G. W., \& Haghighipour, N., Rapid Formation of 
Ice Giant Planets, 2002, Icarus, 156, 291-295.

\bibitem{} 
Brownlee, D. E. et al., Comet 81P/Wild 2 Under a Microscope,
2006, Science, 314, 1711-1716.

\bibitem{} 
Ciesla, F. J., Outward Transport of High-Temperature Materials
Around the Midplane of the Solar Nebula, 2007, Science, 318, 613-615.

\bibitem{} 
Ciesla, F. J., \& Cuzzi, J. N., The evolution of the water vapor
distribution in a viscous protoplanetary disk, 2006, Icarus, 181, 178-204.

\bibitem{} 
Cieza, L., et al., The {\it SPITZER} c2d Survey of Weak-Line T Tauri
Stars. II. Astrophysical Journal, 667, 308-328.

\bibitem{} 
Clayton, R. N., Oxygen Isotopes in Meteorites, 1993, Annual Reviews 
of Earth Planetary Science, 21, 115-149.

\bibitem{} 
Clayton, R. N., Self-shielding in the solar nebula, 2002, Nature,
415, 860-861.

\bibitem{} 
Desch, S. J., \& Connolly, H. C., Jr., A model of the thermal
processing of particles in solar nebula shocks: Application to
the cooling rates of chondrules, 2002,
Meteoritics \& Planetary Science, 37, 183-207.

\bibitem{} 
Gail, H.-P., Radial mixing in protoplanetary disks III. Carbon dust 
oxidation and abundance of hydrocarbons in comets, 2002, Astronomy \& 
Astrophysics, 390, 253-265. 

\bibitem{} 
Halliday, A., The clock's second hand, 2004, Nature, 431, 
253-254. 

\bibitem{} 
Hsu, W., Huss, G. R., \& Wasserburg, G. J., Al-Mg systematics 
of CAIs, POI, and ferromagnesian chondrules from Ningqiang, 2003,
Meteoritics \& Planetary Science, 38, 35-48.

\bibitem{} 
Inaba, S., Wetherill, G. W., \& Ikoma, M., Formation of gas giant
planets,  core accretion models with fragmentation and planetary
envelope, 2003, Icarus, 166, 46-62.

\bibitem{} 
Kitamura, Y., et al., Investigation of the Physical Properties of
Protoplanetary Disks Around T Tauri Stars by a 1 Arcsecond Imaging
Survey, 2002, Astrophysical Journal, 581, 357-380.

\bibitem{} 
Krot, A. N., Yurimoto, H., Hutcheon, I. D., \& MacPherson, G. J.,
Chronology of the early Solar System from chondrule-bearing
calcium-aluminium-rich inclusions, 2005,
Nature, 434, 998-1001. 

\bibitem{} 
Krot, A. N., et al., Evolution of Oxygen Isotopic Composition 
in the Inner Solar Nebula, 2005, Astrophysical Journal, 622, 1333-1342.

\bibitem{} 
Lyons, J. R., \& Young, E. D., CO self-shielding as the origin of 
oxygen isotope anomalies in the early solar nebula, 2005,
Nature, 435, 317-320. 

\bibitem{} 
Matsumura, S., \& Pudritz, R. E., Dead zones and extrasolar planetary 
properties, 2006, Monthly Notices Royal Astronomical Society, 365, 572-584. 

\bibitem{} 
Mer\'in, B., et al., Abundant Crystalline Silicates in the Disk of a Very
Low Mass Star, 2007, Astrophysical Journal, 661, 361-367.

\bibitem{}
Mostefaoui, S., Lugmair, G., \& Hoppe, P., $^{60}$Fe,  A Heat 
Source for Planetary Differentiation From a Nearby Supernova Explosion, 
2005, Astrophysical Journal, 625, 271-277.

\bibitem{} 
Nuth, J. A., \& Johnson, N. M., Crystalline silicates in comets, 
How did they form?, 2006, Icarus, 180, 243-250.

\bibitem{} 
Sakamoto, N., Seto, Y., Itoh, S., Kuramoto, K., Fujino, K., Nagashima,
K., Krot, A.  N., \& Yurimoto, H., Remants of the Early Solar System
Water Enriched in Heavy Oxygen Isotopes, 2007, Science, 317, 231-233.

\bibitem{} 
Simon, S. B., Davis, A. M., Grossman, L., \& Zinner, E. K., 
Origin of hibonite-pyroxene spherules found in carbonaceous chondrites,
1998, Meteoritics \& Planetary Science, 33, 411-424. 

\bibitem{} 
Shu, F. H., Shang, H., Gounelle, M., Glassgold, A. E., \& Lee, T., 
The Origin of Chondrules and Refractory Inclusions in Chondritic 
Meteorites, 2001, Astrophysical Journal, 548, 1029-1050. 

\bibitem{} 
Tachibana, S., \& Huss, G. R., The Initial Abundance of $^{60}$Fe in 
the Solar System, 2003, Astrophysical Journal, 588, L41-L44. 

\bibitem{} 
Thrane, K., Bizzarro, M., \& Baker, J. A., Extremely Brief Formation
Interval for Refractory Inclusions and Uniform Distribution of
$^{26}$Al in the Early Solar System, 2006, Astrophysical Journal,
646, L159-L162.

\bibitem{} 
Tscharnuter, W. M., \& Gail, H.-P., 2-D preplanetary accretion disks I. 
Hydrodynamics, chemistry, and mixing processes, 2007, Astronomy \&
Astrophysics, 463, 369-392.

\bibitem{} 
van Boekel, R., et al., A 10 $\mu$m spectroscopic survey of Herbig
Ae star disks: Grain growth and crystallization, 2005, Astronomy \&
Astrophysics, 437, 189-208.

\bibitem{}
Vanhala, H. A. T., \& Boss, A. P., Injection of Radioactivities into
the Forming Solar System, 2002, Astrophysical Journal, 575, 1144-1150.

\bibitem{} 
Weidenschilling, S. J., Formation processes and time scales for 
meteorite parent bodies, 1988, in Meteorites and the Early Solar System,
eds. J. F. Kerridge \& M. S. Matthews (Tucson,  Univ. of Arizona Press), 
pp. 348-371. 
 
\bibitem{} 
Yin, Q.-Z., Predicting the Sun's Oxygen Isotope Composition, 2004, 
Science, 305, 1729-1730.

\bibitem{} 
Yokoyama, T., Rai, V. K., Alexander, C. M. O'D., Lewis, R. S.,
Carlson, R. W., Shirey, S. B., Thiemens, M. H., \& Walker, R. J.,
2007, Osmium isotope evidence for uniform distribution of $s-$
and $r-$process components in the early solar system, Earth \&
Planetary Science Letters, 259, 567-580.

\bibitem{} 
Young, E. D., et al., Supra-Canonical $^{26}$Al/$^{27}$Al and the
Residence Time of CAIs in the Solar Protoplanetary Disk, 2005,
Science, 308, 223-227.

\bibitem{}
Yurimoto, H., \& Kuramoto, K., Molecular Cloud Origin for the 
Oxygen Isotope Heterogeneity in the Solar System, 2004, 
Science, 305, 1763-1766.

\end{thebibliography}
\end{document}